\documentclass[10pt,preprint2]{aastex}

\usepackage{graphicx}
\usepackage{longtable}
\begin{document}

\title{The W~UMa binaries
USNO-A2.0~1350-17365531, V471~Cas, V479~Lac and V560~Lac: light
curve solutions and global parameters based on the GAIA
distances}


\author{Diana~P.~Kjurkchieva\altaffilmark{1}, Velimir ~A.~Popov\altaffilmark{1}, Yordanka Eneva\altaffilmark{1,2} and Nikola~I.~Petrov\altaffilmark{3}}
\altaffiltext{1}{Department of Physics and Astronomy, Shumen
University, 9700 Shumen, Bulgaria} \altaffiltext{2}{Medical
University, Varna, 84 Tcar Osvoboditel str., Bulgaria}
\altaffiltext{3}{Institute of Astronomy and NAO, Bulgarian Academy
of Sciences, 72 Tsarigradsko Shose Blvd., 1784 Sofia, Bulgaria}

\begin{abstract}
We present photometric observations in Sloan filters $g', i'$ of
the eclipsing W UMa stars USNO-A2.0~1350-17365531, V471 Cas, V479
Lac and V560 Lac. The sinusoidal-like O-C diagram of V471
Cas shows presence of third body with mass of 0.12 $M_{\odot}$
(red dwarf) at distance 897 $R_{\odot}$. The O-C diagram of V479
Lac reveals period decreasing of $dP/dt$ = –-1.69 $\times$
10$^{-6}$ d yr$^{-1}$. The results of the light curve solutions
are: (i) The targets are overcontact binaries with small fill-out
factors; (ii) Their components are F--K stars, comparable
in size, whose temperature differences are below 80 K; (iii) All
targets undergo partial eclipses and to limit the possible mass
ratios we carried out two-step $q$-search analysis. The target
global parameters (luminosities, radii, masses) were obtained on
the base of their GAIA distances and the results of our light
curve solutions. The obtained total mass of V560 Lac turns
out smaller than the lower mass limit for the presently existing W
UMa binaries of 1.0 -- 1.2 $M_{\odot}$, i.e. this target is an
peculiar overcontact system.
\end{abstract}


\section{INTRODUCTION}

The high-precision positions in the Hertzsprung-Russell diagram of
stars of known surface abundances, provided by Hipparcos and by
high-resolution spectroscopy, revealed discrepancies between the
observations and the predictions of standard stellar models
(Perryman et al. 1995). Hence, although the fundamental principles
of stellar evolution are well known, there are some aspects of the
evolution and stellar interior which require further improvement
of the theories. These problems need precise fundamental
parameters of stars in different stages of their evolution.

Eclipsing binary systems are the most important sources of such
information, especially the numerous W UMa-type binaries
consisting of two main sequence stars embedded by a common
convective photosphere (Lucy 1968a, 1968b).

The determination of the global parameters of W UMa systems is
difficult because the photometric mass ratio of the most of them
which undergo partial eclipses is poorly estimated (Rucinski 2001,
Terrell $\&$ Wilson 2005). Moreover, their spectral mass ratios
are not precise due to the highly broadened and blended spectral
lines of the components (Frasca et al. 2000; Bilir et al. 2005;
Dall $\&$ Schmidtobreick 2005).

The GAIA mission (GAIA collaboration 2018) opens new horizons in
the study of the W UMa stars because provides unprecedented
parallax measurements of about one billion stars in our Galaxy.
They would allow high-precision determination of the global
parameters of many ground-based observed eclipsing binaries.

This paper presents photometric observations of the short-period W
UMa-type systems USNO-A2.0 1350-17365531 (further on assigned as
USNO 1350), V471 Cas, V479 Lac and V560 Lac. Table 1 presents
information for their coordinates and variability from VSX
database. The goal of our study was to determine their parameters
by light curve solutions and GAIA distances as well as to search
for period changes.

\begin{table*}[ht!]
\begin{scriptsize}
\begin{center}
\caption{Parameters of the targets from the VSX database.}
\label{tab:1}
\begin{tabular}{ccllllll}
\hline
Target          &   RA          &   DEC       &   mag      & ampl  &   P        &  type & Ref   \\
\hline
USNO 1350       &   22 48 35.91 & +50 49 06.4 & 14.48(CR)  & 0.48  &   0.26638  &   EW   & Lapukhin et al. 2014 \\
V471 Cas        &   01 32 20.48 & +55 12 19.8 & 14.1(p)    & 0.5   &   0.335998 & EW/KW  & Hoffmeister 1966 \\
V479 Lac        &   22 52 50.69 & +35 58 56.5 & 12.2(R)    & 0.70  &   0.34575  &   EW   & Khruslov 2008 \\
V560 Lac        &   22 48 32.56 & +50 49 35.4 & 14.34(CR)  & 0.64  &   0.2722   &   EW   & Lapukhin et al. 2013\\
\hline
\end{tabular}
\end{center}
\end{scriptsize}
\end{table*}

\begin{table*}[ht!]
\begin{scriptsize}
\begin{center}
\caption{Journal of our photometric observations.} \label{tab:2}
\begin{tabular}{clccc}
\hline
Star                &  Date       & Exposures& Number & error \\
 \hline
USNO 1350           & 2017 Oct 9  & 240, 240 & 21, 21 & 0.017, 0.014 \\
                    & 2017 Oct 10 & 240, 240 & 37, 37 & 0.030, 0.026 \\
                    & 2017 Oct 11 & 240, 240 & 41, 41 & 0.011, 0.012 \\
\hline
V471 Cas            & 2017 Nov 22 & 150, 240 & 86, 86 & 0.006, 0.007 \\
                    & 2017 Nov 23 & 150, 240 & 81, 81 & 0.005, 0.006 \\
\hline
V479 Lac            & 2017 Oct 12 & 60, 120 & 140, 140 & 0.005, 0.005 \\
                    & 2017 Oct 13 & 60, 120 & 62, 62   & 0.006, 0.007 \\
                    & 2017 Oct 14 & 60, 120 & 137, 137 & 0.010, 0.010 \\
\hline
V560 Lac            & 2017 Oct 9  & 240, 240 & 21, 21 & 0.014, 0.013 \\
                    & 2017 Oct 10 & 240, 240 & 39, 39 & 0.022, 0.027 \\
                    & 2017 Oct 11 & 240, 240 & 41, 41 & 0.009, 0.012 \\
\hline
\end{tabular}
\end{center}
\end{scriptsize}
\end{table*}

\section{OBSERVATIONS}

Our CCD photometric observations of the targets in Sloan $g', i'$
bands were carried out with the 30-cm Ritchey Chretien Astrograph
(located into the \emph{IRIDA South} dome) using CCD camera ATIK
4000M. Information for our observations is presented in Table~2.

The photometric data were reduced by {\textsc{AIP4WIN2.0} (Berry
$\&$ Burnell 2005). An aperture ensemble photometry was performed
with the software \textsc{VPHOT} using more than six standard
stars (Table 3) in the observed field whose coordinates were taken
from the catalogue UCAC4 (Zacharias et al. 2013) and their
magnitudes from the catalogue APASS DR9.

\begin{table*}[ht!]
\begin{scriptsize}
\begin{center}
\caption{List of standard stars.} \label{tab:3}
\begin{tabular}{ccllll}
  \hline
Label & Star ID       &  RA         & Dec          & $g'$   & $i'$ \\
 \hline
Target & USNO 1350    & 22 48 35.91 & +50 49 06.4  & 15.17  & 13.91 \\
Chk & UCAC4 705-109759& 22 48 35.25 & +50 49 45.32 & 14.260 & 13.042 \\
C1 & UCAC4 705-109766 & 22 48 37.22 & +50 53 39.86 & 12.768 & 11.229 \\
C2 & UCAC4 705-109796 & 22 48 46.35 & +50 48 49.30 & 13.931 & 13.047 \\
C3 & UCAC4 705-109827 & 22 48 55.60 & +50 49 38.22 & 13.099 & 12.509 \\
C4 & UCAC4 705-109708 & 22 48 19.49 & +50 49 49.20 & 13.885 & 13.275 \\
C5 & UCAC4 705-109721 & 22 48 23.35 & +50 49 22.84 & 14.237 & 13.746 \\
C6 & UCAC4 704-108193 & 22 48 01.25 & +50 47 31.75 & 13.654 & 13.005 \\
C7 & UCAC4 704-108347 & 22 48 40.68 & +50 47 38.84 & 13.185 & 12.562 \\
C8 & UCAC4 705-109878 & 22 49 09.62 & +50 48 21.86 & 12.863 & 12.240 \\
\hline
Target & V471 Cas     & 01 32 20.48 & +55 12 19.8  & 13.87  & 13.07 \\
Chk & UCAC4 727-012640& 01 32 11.74 & +55 13 34.01 & 14.255 & 13.505 \\
C1 & UCAC4 727-012714 & 01 32 39.00 & +55 15 08.04 & 13.768 & 13.253 \\
C2 & UCAC4 727-012652 & 01 32 17.81 & +55 14 52.41 & 13.096 & 12.673 \\
C3 & UCAC4 727-012622 & 01 32 02.18 & +55 14 33.01 & 13.107 & 12.420 \\
C4 & UCAC4 727-012552 & 01 31 37.73 & +55 14 17.12 & 13.275 & 12.682 \\
C5 & UCAC4 726-012990 & 01 32 44.12 & +55 10 24.72 & 14.484 & 12.894 \\
C6 & UCAC4 726-012935 & 01 32 18.10 & +55 09 03.67 & 14.092 & 12.411 \\
C7 & UCAC4 727-012662 & 01 32 21.12 & +55 18 24.56 & 14.185 & 11.800 \\
C8 & UCAC4 727-012588 & 01 31 49.76 & +55 13 56.00 & 13.789 & 13.229 \\
C9 & UCAC4 727-012623 & 01 32 02.28 & +55 13 20.10 & 12.316 & 11.625 \\
C10& UCAC4 726-012820 & 01 31 40.78 & +55 11 26.78 & 14.233 & 13.466 \\
\hline
Target & V479 Lac     & 22 52 50.69 & +35 58 56.5  & 12.49  & 11.75 \\
Chk & UCAC4 630-129893& 22 52 59.78 & +35 58 32.07 & 13.770 & 12.215 \\
C1 & UCAC4 630-129875 & 22 52 43.78 & +35 53 55.03 & 12.034 & 11.603 \\
C2 & UCAC4 630-129912 & 22 53 19.06 & +35 56 53.76 & 13.762 & 13.257 \\
C3 & UCAC4 630-129921 & 22 53 26.62 & +35 59 53.33 & 13.413 & 12.547 \\
C4 & UCAC4 631-132559 & 22 52 30.81 & +36 02 51.51 & 12.757 & 12.171 \\
C5 & UCAC4 631-132585 & 22 52 54.16 & +36 00 31.90 & 11.904 & 11.351 \\
C6 & UCAC4 631-132557 & 22 52 29.70 & +36 00 15.27 & 13.990 & 13.333 \\
\hline
Target & V560 Lac     & 22 48 32.56 & +50 49 35.4  & 14.90  & 13.89 \\
Chk &UCAC4 705-109759 & 22 48 35.25 & +50 49 45.32 & 14.260 & 13.042 \\
C1 & UCAC4 705-109766 & 22 48 37.22 & +50 53 39.86 & 12.768 & 11.229 \\
C2 & UCAC4 705-109796 & 22 48 46.35 & +50 48 49.30 & 13.931 & 13.047 \\
C3 & UCAC4 705-109827 & 22 48 55.60 & +50 49 38.22 & 13.099 & 12.509 \\
C4 & UCAC4 705-109708 & 22 48 19.49 & +50 49 49.20 & 13.885 & 13.275 \\
C5 & UCAC4 705-109721 & 22 48 23.35 & +50 49 22.84 & 14.237 & 13.746 \\
C6 & UCAC4 704-108193 & 22 48 01.25 & +50 47 31.75 & 13.654 & 13.005 \\
C7 & UCAC4 704-108347 & 22 48 40.68 & +50 47 38.84 & 13.185 & 12.562 \\
C8 & UCAC4 705-109878 & 22 49 09.62 & +50 48 21.86 & 12.863 & 12.240 \\
\hline
\end{tabular}
\end{center}
\end{scriptsize}
\end{table*}

\begin{figure*}
\centerline{\includegraphics[width=15cm]{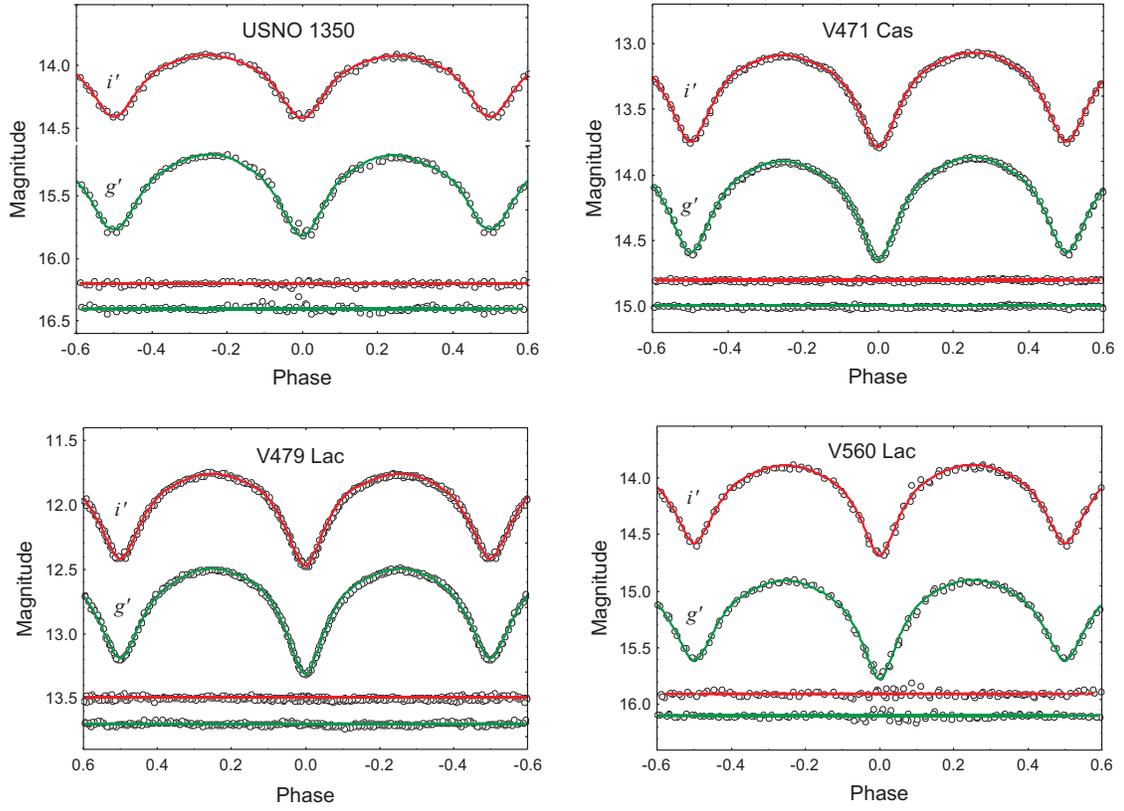}}
\caption{Top of each panel: the folded light curves and their
fits; Bottom: the corresponding residuals (shifted vertically by
different amount to save space). \label{Fig1}}
\end{figure*}

\section{LIGHT CURVE SOLUTIONS}

We carried out the modeling of our data by the package
\textsc{PHOEBE} (Prsa $\&$ Zwitter 2005, Prsa et al. 2011, 2016)
which is based on the Wilson–-Devinney (WD) code (Wilson $\&$
Devinney 1971, Wilson 1979, 1993). It allows simultaneous modeling
of photometric data in a number of filters and provides a
graphical user interface.

The observational data (Fig. 1) show that the targets are
overcontact systems and we modelled them using the corresponding
mode ''Overcontact binary not in thermal contact'' of
\textsc{PHOEBE}.

Table 4 presents the target temperatures $T_{m}^{ci}$
determined by the observed dereddened color indices
$(g'-i')$ at quadratures and the relation of Covey et al. (2007),
as well as their temperatures $T_{m}^{G}$ from GAIA DR2
(GAIA collaboration 2018) and $T_{m}^{L}$ from LAMOST
(Luo et al. 2015). The differences in the temperature values may
due, at least partially, to the different (inappropriate) phase of
measurements of GAIA and LAMOST. The last column reveals the
adopted {$T_{m}$ values in the procedure of the light
curve solution.

\begin{table}[ht!]
\begin{scriptsize}
\begin{center}
\caption{Target temperatures.} \label{tab:4}
\begin{tabular}{cclll}
\hline
Target        & $T_{m}^{ci}$ & $T_{m}^{G}$ & $T_{m}^{L}$ & $T_{m}$     \\
   \hline
USNO 1350       &   5140    & 5095     & -          & 5140   \\
V471 Cas        &   6000    & 5347     & -          & 6000   \\
V479 Lac        &   5680    & 5361     & 6220       & 5680   \\
V560 Lac        &   5650    & 5000     & -          & 5650   \\
\hline
\end{tabular}
\end{center}
\end{scriptsize}
\end{table}

We fixed the primary temperature $T_{1}$ = $T_{m}$ and searched
for the best fit varying initial epoch $T_0$, period $P$,
secondary temperature $T_{2}$, mass ratio $q$, inclination $i$ and
potential $\Omega$. We adopted coefficients of gravity brightening
0.32 and reflection effect 0.5 appropriate for late stars (Table
4). The limb-darkening coefficients were interpolated according to
the tables of Van Hamme (1993). In order to reproduce the light
curve distortions we used cool spots and varied their parameters
(longitude $\lambda$, latitude $\beta$, angular size $\alpha$ and
temperature factor $\kappa$). Hence, each spot introduce 4
new parameters which cannot be unambiguously determined (excepting
spot longitude). That is why we used the following considerations
to reduce the spot parameters: (i) Due to lack of additional
knowledge (for instance spectral or polarimetric data) the spots
were put on the primary (hotter) component but the same result
could be reached by spot on the secondary component with bigger
temperature contrast; (ii) We used equatorial spots, i.e. the spot
latitude was fixed; (iii) The temperature contrast $\kappa =
(T_{ph}-T_{sp})/T_{ph}$ was varied in the range 0.8--0.9
appropriate for the component temperature (Berdyugina 2005).

The eclipses of all targets do not contain flat bottom (Fig. 1)
that means partial eclipses. Hence, their photometric mass ratios
are poorly determined (Rucinski 2001, Terrell $\&$ Wilson 2005)
and require $q$-search analysis. Firstly, we varied the mass ratio
in a wide interval, from 0.1 to 10.0, to obtain the global minimum
of the $q$-search curves (Fig. 2). In order to limit further the
possible mass ratios we mapped the $\chi^2$ dependence on $q$ (for
values within the global minimum) and orbital inclination $i$
(Fig. 3). The obtained values of $q$ and $i$ were used in the last
stage of the light curve solution.

\begin{figure*}
\centerline{\includegraphics[width=12cm]{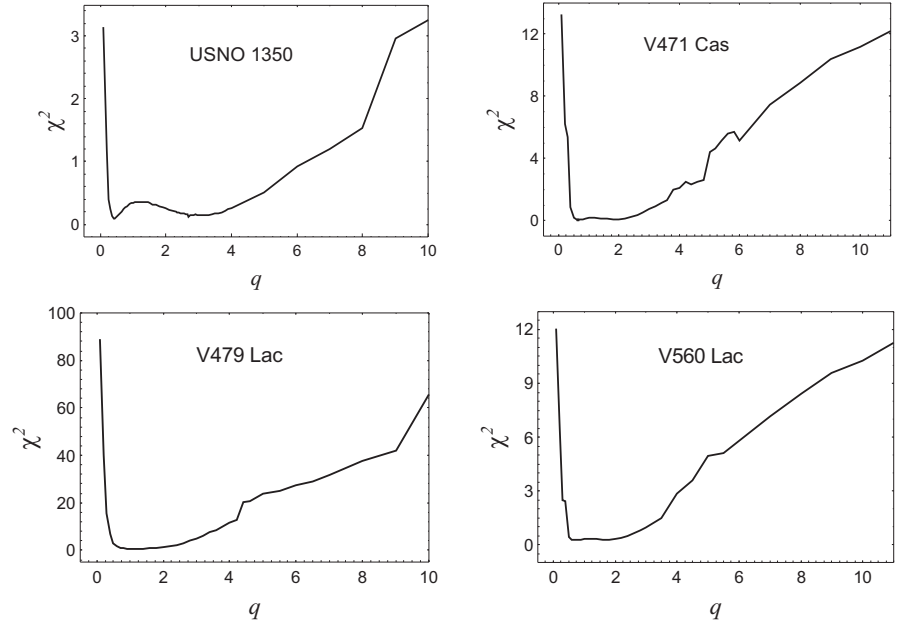}}
\caption{$q$-search curves for the four targets. \label{Fig2}}
\end{figure*}

After reaching the best light curve solution we adjusted the
stellar temperatures $T_{1}$ and $T_{2}$ around the value $T_m$ by
the formulae (Kjurkchieva $\&$ Vasileva 2015)
\begin{equation}
T_{1}^{f}=T_{\rm {m}} + \frac{c \Delta T}{c+1}; \quad
T_{2}^{f}=T_{1}^{f}-\Delta T
\end{equation}
where the quantities $c=l_2/l_1$ (the ratio of the relative
luminosities of the stellar components) and $\Delta T=T_{m}-T_{2}$
are determined from the \textsc{PHOEBE} solution.

Last fitting procedure was carried out for fixed
$T_{1}^{f}$ and $T_{2}^{f}$ and corresponding limb-darkening
coefficients in order to obtain the final and self-consistent
solution. 

\begin{figure*}
\centerline{\includegraphics[width=16cm]{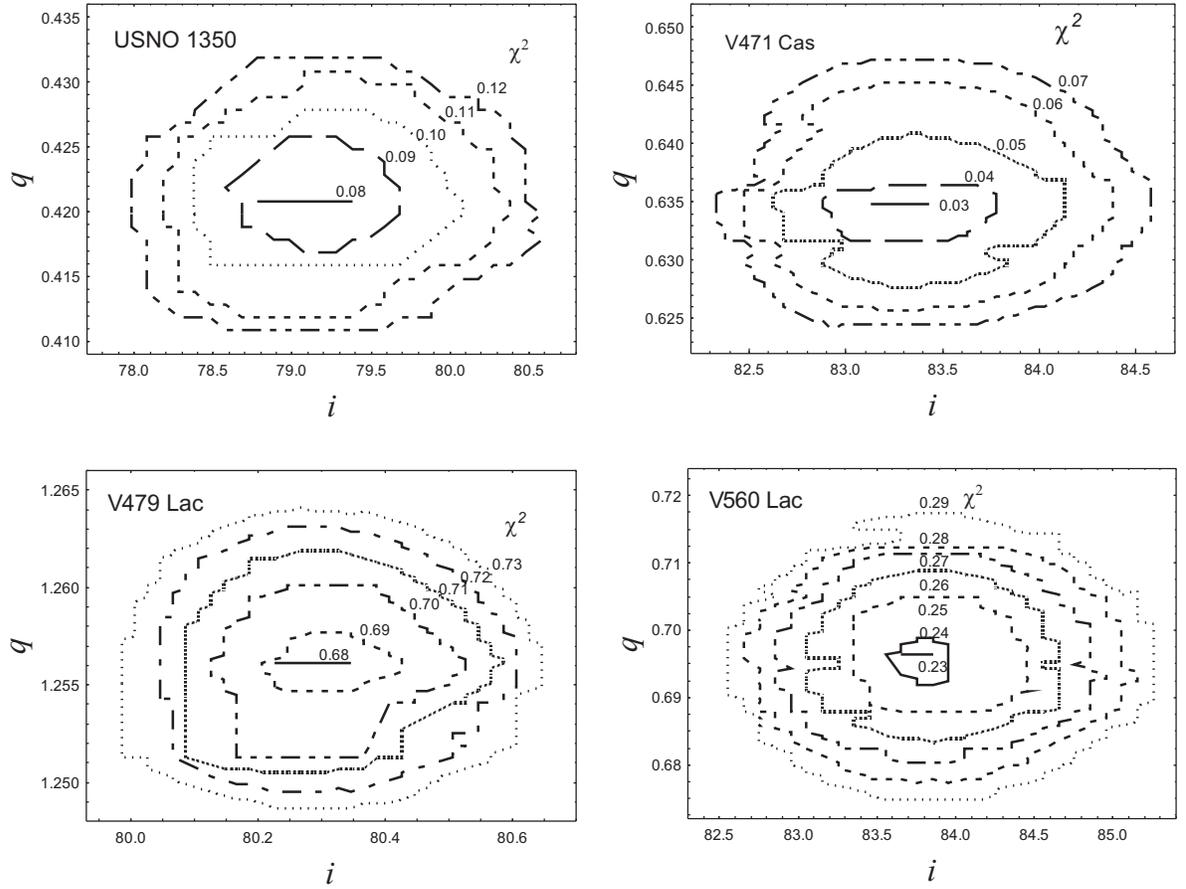}} \caption{The
$\chi^2$ dependencies on mass ratio $q$ and inclination $i$: the
different isolines circumscribe the areas whose normalized
$\chi^2$ are smaller than the marked values; the circle
corresponds to the minimum of $\chi^2$. \label{Fig3}}
\end{figure*}

\begin{figure}
\centerline{\includegraphics[width=8cm]{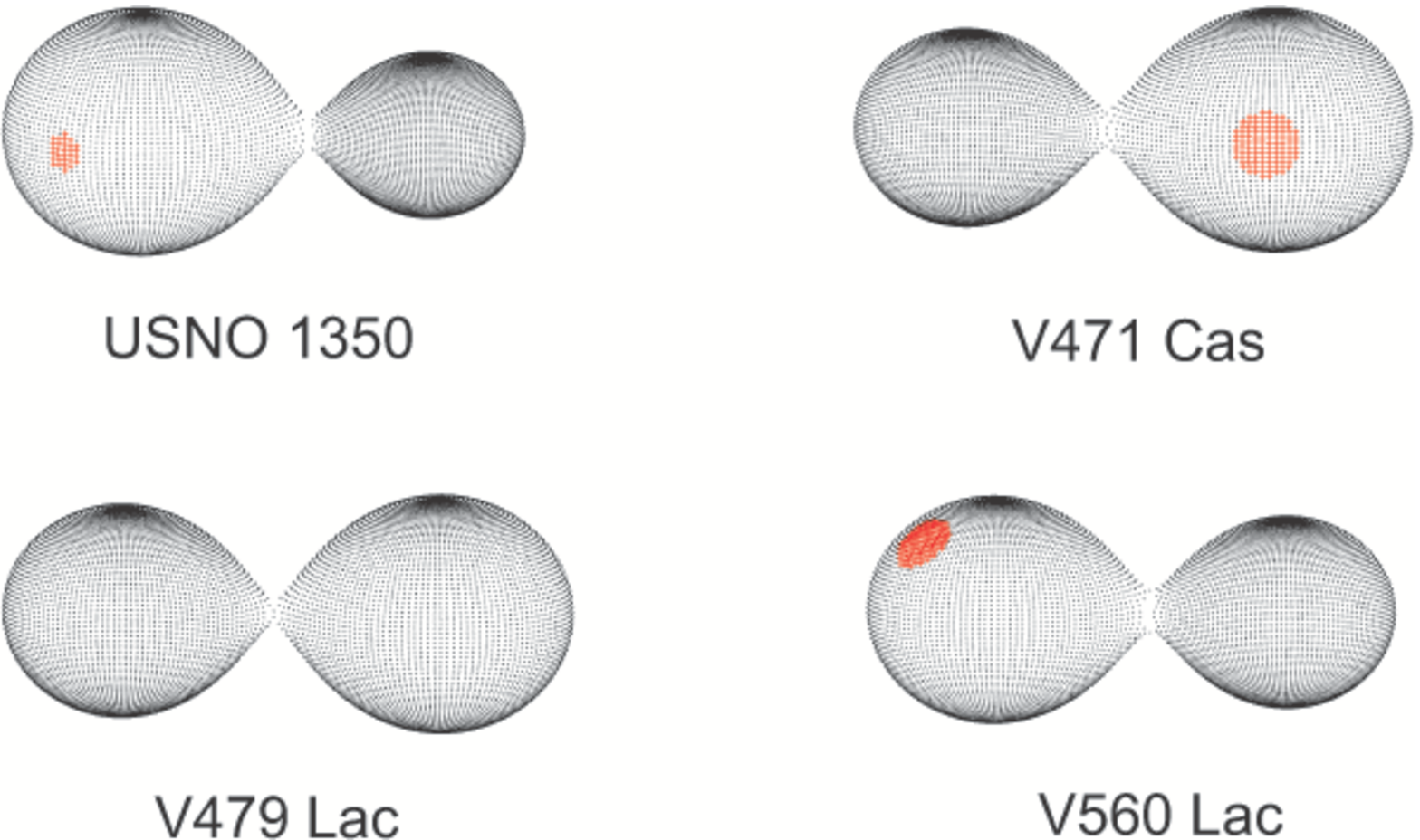}}
\caption{Three-dimensional (3D) configurations of the targets made
using Binary Maker 3 by Bradstreet $\&$ Steelman (2002).
\label{Fig4}}
\end{figure}

Table~5 contains the final values of the fitted stellar parameters
and their uncertainties while Table 6 exhibits the calculated
parameters: relative stellar radii $r_{1, 2}$; fill-out factor
$f$; ratio of relative stellar luminosities $l_2/l_1$. Their
errors are determined from the uncertainties of fitted parameters
used for their calculation.

The synthetic curves corresponding to the parameters of our light
curve solutions are shown in Fig. 1 as continuous lines while Fig.
4 exhibits the target three-dimensional configurations.

\begin{table*}[ht!]
\begin{scriptsize}
\begin{center}
\caption{Fitted parameters of the best light curve solutions}
\label{tab:5}
\begin{tabular}{cllllllllll}
  \hline
Star         & $T_0$                & $P$         &  $i$     &  $q$    & $T_{2}$  & $\Omega$  & $\beta$ & $\lambda$ & $\alpha$ & $\kappa$   \\
             & [d]                  & [d]         & [$^{\circ}$] &     & [K]  &    & [$^{\circ}$] & [$^{\circ}$] & [$^{\circ}$] &    \\
 \hline
USNO 1350    & 8037.514889(0.00024) & 0.2663764   & 79.08(1.7) & 0.421(0.07)& 5090(40) & 2.70(0.2)  &   90    &   238(1)  &   8(1)   &   0.89(1)       \\
V471 Cas     & 8080.254983(0.0004)  & 0.40093526  & 83.28(0.5) & 0.635(0.04)& 5940(30) & 3.098(0.07)&   90    &   90(1)   &   15(1)  &   0.85(1) \\
V479 Lac     & 8039.299724(0.00012) & 0.3457586(3)& 80.37(0.5) & 1.256(0.04)& 5620(20) & 4.141(0.05)&         &           &          &       \\
V560 Lac     & 8038.60221(0.00012)  & 0.2722467(3)& 82.03(2.4) & 0.697(0.09)& 5575(40) & 3.199(0.2) &   50    &   220(1)  &   15(1)  &   0.90(1) \\
   \hline
\end{tabular}
\end{center}
\end{scriptsize}
\end{table*}

\begin{table*}[ht!]
\begin{scriptsize}
\begin{center}
\caption{Calculated parameters} \label{tab:6}
\begin{tabular}{cclllll}
  \hline
Star            & $T_{1}^{f}$ & $T_{2}^{f}$  & $r_{1}$   & $r_{2}$   & $f$    & $l_{2}/l_{1}$ \\
 \hline
USNO 1350       &   5155(38)  &   5104(91)   & 0.462(26) & 0.312(24) & 0.077  &   0.436   \\
V471 Cas        &   5975(29)  &   6035(69)   & 0.427(6)  & 0.347(9)  & 0.078  &   0.692   \\
V479 Lac        &   5713(22)  &   5652(19)   & 0.362(9)  & 0.402(9)  & 0.026  &   1.180   \\
V560 Lac        &   5680(22)  &   5604(19)   & 0.420(9)  & 0.356(9)  & 0.095  &   0.681   \\
   \hline
\end{tabular}
\end{center}
\end{scriptsize}
\end{table*}

\begin{table*}[ht!]
\begin{scriptsize}
\begin{center}
\caption{Global parameters.} \label{tab:7}
\begin{tabular}{ccllccclllll}
  \hline
Target          &  $d$ &$M_{bol}$& $L$      & $L_1$    & $L_2$    & $R_1$ & $R_2$ & $a$ & $M$  & $M_1$    & $M_2$  \\
                &  [pc]&         &[$L_{\odot}$]&[$L_{\odot}$]&[$L_{\odot}$]&[$R_{\odot}$]&[$R_{\odot}$]&[$R_{\odot}$]&[$M_{\odot}$]&[$M_{\odot}$]& [$M_{\odot}$]\\
   \hline
USNO 1350       &   629 & 5.177   & 0.681     &  0.474   &  0.207   & 0.866 & 0.583 & 1.86 & 1.233 & 0.868 & 0.365 \\
V471 Cas        &   754 & 4.063   & 1.900     &  1.123   &  0.777   & 0.992 & 0.808 & 2.33 & 1.054 & 0.645 & 0.410  \\
V479 Lac        &   370 & 4.159   & 1.739     &  0.798   &  0.942   & 0.914 & 1.015 & 2.52 & 1.803 & 0.799 & 1.004 \\
V560 Lac        &   667 & 5.174   & 0.683     &  0.405   &  0.277   & 0.660 & 0.560 & 1.57 & 0.703 & 0.414 & 0.289 \\
   \hline
\end{tabular}
\end{center}
\end{scriptsize}
\end{table*}

\begin{figure}
\centerline{\includegraphics[width=8cm]{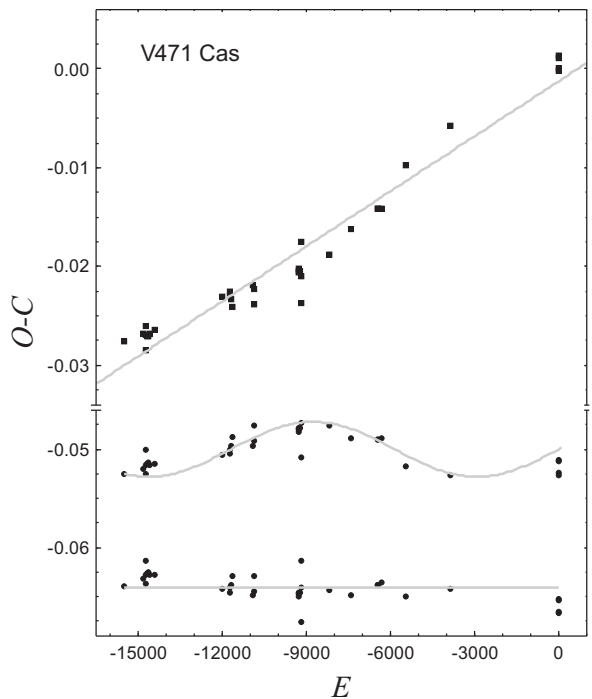}} \caption{Top
panel: O-C data of V471 Cas and their linear fit; Middle panel: O-C data minus linear fit (shifted vertically);
Bottom panel: residuals (O-C data minus linear and sinusoidal fit, shifted vertically). \label{Fig5}}
\end{figure}

\begin{figure}
\centerline{\includegraphics[width=8cm]{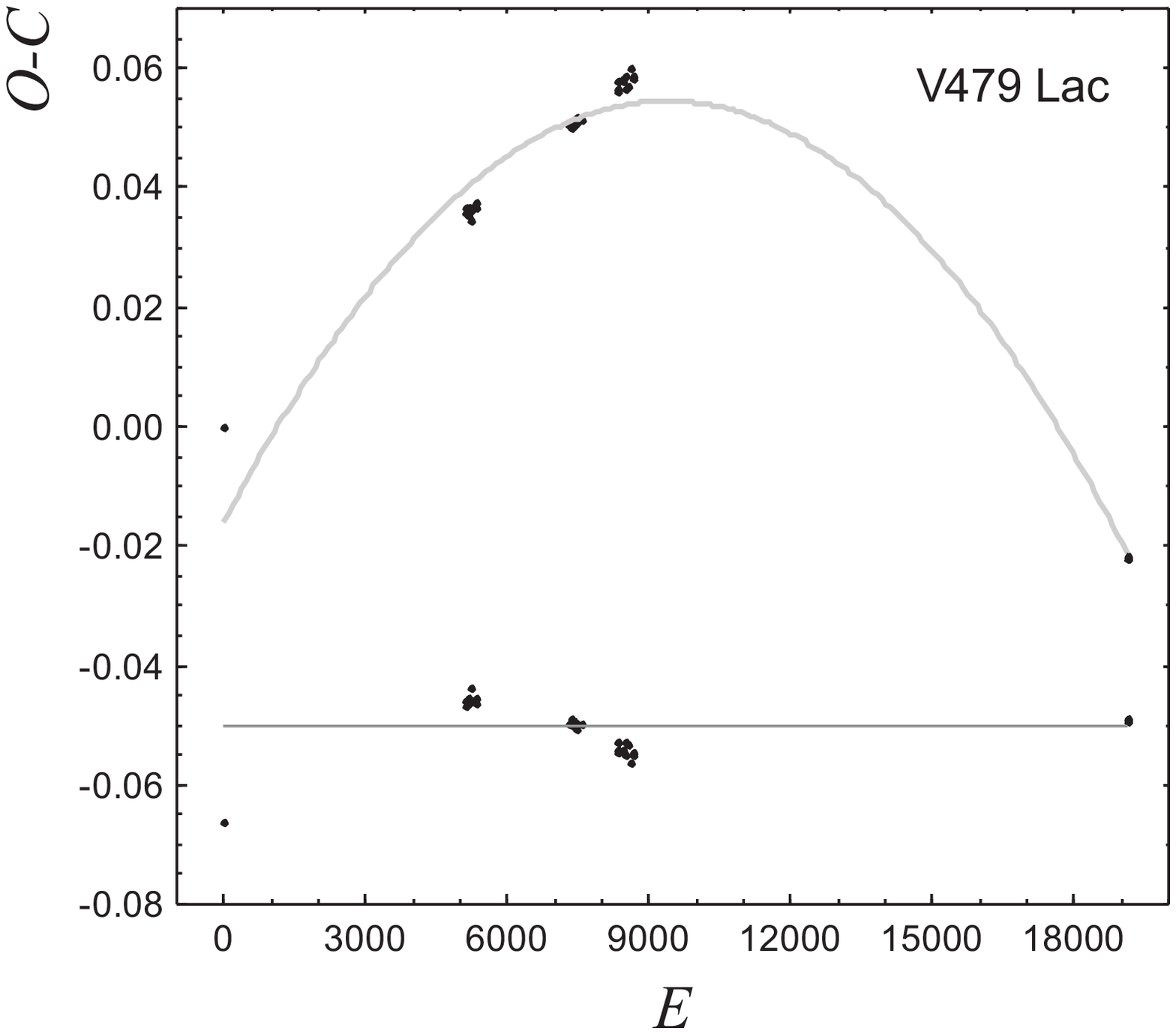}} \caption{Top
panel: O-C data of V479 Lac and their quadratic fit;
Bottom panel: residuals (O-C data minus quadratic fit, shifted vertically). \label{Fig6}}
\end{figure}

\section{Global parameters}

The target global parameters (Table 7) were calculated by the
following procedure.

(1) We determined absolute target magnitude $M_{V}$ from
the GAIA distance $d$ (Bailer-Jones et al. 2018) and target
magnitude $V$ (corrected for extinction) by the formula for the
distance modulus. The absolute target magnitude $M_{bol}$ was
calculated from $M_{V}$ and bolometric correction corresponding to
the target temperature. Then the absolute target luminosity $L$
was obtained.

(2) PHOEBE yields as output parameters bolometric magnitudes of
the two components $M_{bol}^i$ in conditional units (if there is
not radial velocity curves) but their difference
$M_{bol}^2-M_{bol}^1$ determines the true luminosity ratio
$l_2/l_1=L_2/L_1$. Its value and that of $L$ allowed us to
calculate $L_1$ and $L_2$.

(3) The absolute component radii $R_i$ were estimated from their
luminosties $L_i$ and temperatures $T_i$ (adopting black-body
emission).

(4) The absolute and relative component radii were used to obtain
the orbital axis $a$.

(5) The target total masses $M$ were calculated by the third
Kepler law from the period $P$ and orbital axis $a$. Then the
component masses $M_i$ were obtained by $M$ and mass ratio $q$.

Thus, the GAIA distances and the foregoing standard
procedure for determination of global parameters of eclipsing
binaries supersede the all old empirical relations (Rucinski 2004,
Gettel et al. 2006, etc.) applied for this aim.

\section{Analysis of the results and conclusions}

The main results from our study of the W UMa-type binaries USNO
1350, V471 Cas, V479 Lac and V560 Lac are as follows.

(1) Initial epochs for the four targets were determined (Table 5)
and their periods were improved.

(2) Our observations confirmed the conclusion of Liu $\&$ Tan
(1991) that the orbital period of V471 Cas is not 0.335998 days
(Table 1) but around 0.4 days. Liu $\&$ Tan (1991) have
obtained 0.405356(22) d.

(3) Besides our observations we found photometric data of V471 Cas
and V479 Lac from SWASP database (Butters et al. 2010) and
determined their times of minima. Moreover, we used the available
times of minima of V471 Cas published in IBVS (Dvorak 2004;
Hubscher 2005, 2011, 2014; Hubscher et al. 2005, 2006; Nelson
2008; Diethelm 2009, 2010, 2011, 2012).

The O-C diagram of V471 Cas contains 34 times of minima
(Fig. 5) which cover around 17 yrs. We excluded the times of
minima obtained by Liu $\&$ Tan (1991) because their observations
are photoelectric and with low time resolution. The O-C diagram
shows sinusoidal variations superposed on linear increasing. The
linear fit of the O-C diagram allowed us to improve the mean
period of V471 Cas to 0.4009371165 d. The sinusoidal curve is with
period of 12.8 yrs and amplitude of 242 seconds. It implies
presence of third body with mass of 0.12 $M_{\odot}$ (possible red
dwarf) orbiting the binary at distance 897 $R_{\odot}$. Future
observations would refine these parameters. 

The O-C diagram of V479 Lac with 41 times of minima (Fig.
6)  which cover around 18 yrs reveals period decreasing of
$dP/dt$ = –-1.69 $\times$ 10$^{-6}$ d yr$^{-1}$.

(4) The $BV$ light curve solution of V471 Cas by Liu $\&$ Tan
(1991) gives parameter values of V471 Cas: $q$ = 0.595; $i$ =
83.29$^{\circ}$; $T_1$ = 5660 K, $T_2$ = 5636 K; $\Omega$=2.986;
$r_1$ = 0.44; $r_2$ = 0.34. These values are within the errors of
ours excluding component temperatures which both are
lower by around 300 K than ours.

(5) All targets are slightly overcontact binaries with fillout
factors up to 0.1.

(6) All targets undergo partial eclipses. We carried out a
two-step $q$-search analysis to limit their possible mass ratios.

(7) The components of the targets are F--K stars. Their
temperature differences are inconsiderable, below 80 K (Table 6).

(8) The light curve distortions of USNO 1350, V471 Cas, and V560
Lac were reproduced by cool spots on their primaries. Summarizing
the optical, X-ray, UV, IR, and radio observations, Dryomova $\&$
Svechnikov (2006) concluded that the contact W UMa-systems have a
highly variable corona, whose appearance and heating mechanism
presuppose the presence  of a  magnetic field generated by a
dynamo mechanism in differentially rotating convective layers. An
indirect confirmation of the existence of magnetic fields in these
late-type stars is their spotted activity.

(9) The differences of the component temperatures (Table 6) are
quite small to determine the subtype, W or A, of our targets.

(10) The components of all targets are comparable in size (Table
6). This is the reason for the partial eclipses even for their
high orbital inclinations.

(11) The GAIA DR2 distances of the targets as well as the
results of our light curve solutions allowed us to estimate the
global parameters of USNO 1350, V471 Cas, V479 Lac and V560 Lac.

(12) The obtained total mass 0.745 $M_{\odot}$ of V560 Lac
is smaller than the lower mass limit for the presently existing
contact binaries of 1.0 -- 1.2 $M_{\odot}$ (Stepien 2006). This
means considerable mass loss and implies late evolutional stage.
Possible mechanisms could be: (i) prolonged intensive mass loss
from the binary during the phase semidetached system; (ii)
sporadic mass losses during some burst-like events. The low mass
makes V560 Lac an peculiar member of the W UMa binaries. Its future
study may throw additional light on the evolution of these
systems. 

The determined masses, radii and luminosities of the target
components present the main contribution of the paper because they
could be used as tests of stellar models of W UMa-type binaries.

\section*{Acknowledgements}

This work was supported partly by project DN08/20 of Scientific
Foundation of the Bulgarian Ministry of Education and Science as
well as by project RD 08-142 of Shumen University. The research
was made with the support of the private IRIDA Observatory
operated remotely (www.irida-observatory.org).
The authors are very grateful to the anonymous Referee for
the valuable notes and recommendations.

This paper makes use of the SIMBAD database, VizieR service, and
Aladin previewer operated at CDS, Strasbourg, France, and NASA's
Astrophysics Data System Abstract Service.

This work has made use of data from the European Space Agency
(ESA) mission Gaia (https://www.cosmos.esa.int/gaia), processed by
the Gaia Data Processing and Analysis Consortium (DPAC,
https://www.cosmos.esa.int/ web/gaia/dpac/consortium). Funding for
the DPAC has been provided by national institutions, in particular
the institutions participating in the Gaia Multilateral Agreement.

This paper makes use of data from the DR1 of the WASP data
(Butters et al. 2010) as provided by the WASP consortium, and the
computing and storage facilities at the CERIT Scientific Cloud,
reg. no. CZ.1.05/3.2.00/08.0144 which is operated by Masaryk
University, Czech Republic.

\end{document}